\documentclass[referee]{raa}
\usepackage{siunitx}
\usepackage{graphicx,times}
\usepackage{natbib}
\usepackage{ctex}
\usepackage{amssymb,amsmath}
\bibpunct{(}{)}{;}{a}{}{,}
\usepackage[english]{babel}
\usepackage{float} 
\usepackage[pagebackref=true]{hyperref}

\begin{document}

   \title{Herbig AeBe Star AS\,442A as UXOR-type system}

 \volnopage{ {\bf 20XX} Vol.\ {\bf X} No. {\bf XX}, 000--000}
   \setcounter{page}{1}

   \author{N.Z. Ismailov\inst{2,*}, F.S. Huseynova\inst{1}, V. Bak{\i}{\c{s}}\inst{3}, N.S. Dzhalilov\inst{2}, S.A. Alishov\inst{2}, and Sh.K. Ismayilova \inst{2}}
   \institute{Batabat Astrophysical Observatory of Nakhchivan State University, Azerbaijan; \\
        \and
Shamakhy Astrophysical Observatory named after N.Tusi, Azerbaijan,
{\it ismailovnshao@gmail.com} \\
\and
Akdeniz University, Space Sciences \& Technologies Dept., Campus, 07058, Antalya, Turkiye\\
\vs \no
   {\small Received 2026 April 15; accepted 2026 April xx}
}

\abstract{ Detailed studies of the physical characteristics of young intermediate-mass Herbig Ae/Be (HAeBe) stars require extensive sets of homogeneous observational data collected over long periods. In this paper, we present the results of a multi-year study focusing on the faint Herbig Ae/Be star AS 442A. The goal of this study was to investigate the spectral and photometric variability of the star, utilizing long-term archival photometry alongside our own spectroscopic observations. The long-term photometric variability was statistically investigated using approximately 3,500 archival measurements spanning 1984 to 2025. Additionally, we examined the star's spectral variability using data obtained at the Shamakhy Astrophysical Observatory (ShAO) between 2020 and 2024. The three deepest brightness minima were detected in 1990, 2013, and 2023, with amplitudes of approximately ∼ 2 mag in the U-band and up to ∼ 1 mag in other bands. All three observed deep minima were accompanied by smaller brightness dips occurring both before and after the main event. The main light curve and color-magnitude diagrams show the ”blueing” effect, which is characteristic of UXOR stars. A quasi-periodic occurrence of deep brightness minima with a period of P = 3759 ± 35 days was detected for the first time. Our spectral observations were performed before the last deep light fading, which occurred in 2023. We show that the observed spectral variability, specifically the strengthening of emission lines and the reduction in depths and widths of absorption lines, is primarily related to the star’s brightness variations. We have shown for the first time that the main reason for the UXOR-like brightness dip of the HAeBe star AS 442A can be explained by a periodic eclipse in a binary system. A close companion as a late-type secondary star with an orbit semimajor axis (a$\geq$ 7.5 au) likely drives the observed stellar variability.
\keywords{Pre-main sequence stars -- circumstellar disks -- UXORs, individual -- AS 442A}
}

   \authorrunning{N.Z. Ismailov  et al.}    
   \titlerunning{Herbig AeBe Star AS442A as UXOR type system}  
   \maketitle
\section{Introduction}
\label{sect:intro}

The star AS\,442A (Sp. B8--A0, $V\sim11.5$\,mag), located in the NGC\,7000\,/\,IC\,5070 star-forming region, was first identified by~\cite{Merrill1950} as an object with a medium-intensity H$\alpha$ emission line.~\citealt{Miller1951} determined from low-resolution spectra that the star's spectral class was B9 or A0, with a weak emission component in the H$\beta$ line. ~\cite{Finkenzeller1984} were the first to show that the star's spectrum has an equivalent width ($EW$) of the H$\alpha$ emission line of $-23$~\AA, and that the star exhibits an infrared (IR) excess. Consequently, it was included in the catalog of Herbig Ae/Be (HAeBe) stars.

The first estimates of the physical parameters of the star were given in the works of~\cite{Mora2001} and~\cite{Montes2009}. The H$\alpha$ emission line and the D Na I doublet lines in the star's spectrum show strong fluctuations. The authors determined the star's spectral type to be B8, its mass to be 3.5\,M$_{\odot}$, the equivalent width of the H$\alpha$ line to be $EW = -32.7$ \,\AA, and the width at 10\% of the line intensity to be H$\alpha$ $W_{10} = 646$\,km\,s$^{-1}$. The H$\alpha$ line has two emission peaks, one of which is less than half the intensity of the other. In addition, the He I $\lambda$ 5876, D Na I, [O I] $\lambda$ 6300, and 6363\,\AA\ lines were detected in the star's spectrum. ~\cite{Corporon1999} detected periodic radial velocity variations in the MgII 4481 \AA\ doublet as well as in the SiII 6347 and 6371 \AA\ absorption doublet and first detected a possible orbital period $P\sim 64$ days and eccentricity $e\sim 0.2$. They also determined the star's temperature to be \num{11000}\,K, the distance to the star to be \num{826}\,pc, the age to be \num{1.5}\,Myr, and the interstellar reddening coefficient $A_{V} = \num{3.85}$~\citep{Mathew2018}.

\(UBVR\) photometric observations of the star were carried out at the Maidanak Observatory in Uzbekistan, and the resulting data were collected in the Strasbourg CDS archive \citep{grankin2007, herbst1994}. In particular, variations in the star's brightness were studied by \citet{melnikov1993} using individual fragments of this material. The star's maximum brightness was shown to exhibit small-amplitude fluctuations around the mean, but in some years, the brightness declined sharply. The brightness fluctuations were shown to have a periodicity of \num{16.2} days with a low degree of confidence \citep{shevchenko1993}. The existence of moderate-amplitude dips in the star's brightness was noted in the work of \citet{herbst1999}. This pattern of brightness variations in AS\,442A resembles a small, actively variable subgroup of young stars, so-called UXORs. These stars are characterized by sudden brightness drops with an amplitude of up to 2--4 mag.

UXORs demonstrate irregular flux variability, which takes the form of sporadic brightness weakening with amplitudes from 2 to 4 magnitudes in the $V$-band. The typical duration of these brightness minima ranges from a few days to a few weeks (see, e.g. \cite{Grinin2023} and references therein). UXORs sometimes exhibit very long minima (more than a year) (see, e.g., \cite{Rostop2012}), which is explained by variable circumstellar (CS) extinction caused by dust from their protoplanetary disks. The disks of these stars form a small line-of-sight angle, i.e., they are nearly edge-on to the line of sight. The UXOR family includes both low-mass T Tauri stars (TTSs) and intermediate-mass Herbig Ae stars (HAEs).

UXORs also exhibit the following characteristic features: during a deep decline in the star's brightness, the linear polarization increases to several percent; as the brightness fades, the color reddens, but after a certain minimum brightness, a "blueing" effect is observed (see, e.g., \citep{Grinin2000}).

The first interpretation of the UXOR-type star CQ Tau was based on the assumption that the system is a binary containing a faint blue companion. Under this scenario, the emission from the secondary component becomes dominant when the primary is obscured by a circumstellar (CS) dust cloud \cite{Wenzel1969}.

One of modern theories attributes these brightness variations to the dimming of the central star by dust clouds orbiting it in a disk located nearly edge-on to the observer \cite{Grinin1994, Natta2000}. When a dust cloud obscures the direct light of the star, the scattered light from the disk becomes the dominant source of radiation for the observer. The formation mechanisms and location of the obscuring dust clouds remain open questions. While the physical explanation for the weakening flux is linked to dust fragments passing in front of the star and eclipsing it, the cause of these sporadic appearances of dust clouds is less clearly established.

Recently, several studies have collected the physical parameters of the star AS\,442A (see, for example, \cite{Vioque2023, Guzman-Diaz2021}). AS\,442A has been shown to form a visual binary system with the component AS\,442B with magnitude \(V \sim \num{15.8}\)\, mag and spectral type G5/K0e \cite{Cohen1979}, which is located at a distance of \num{4.75}\arcsec from the bright central star AS\,442A \cite{Thomas2023}.

Table~\ref{tab1} contains some parameters of the star collected from the literature. The columns list the spectral classes, effective temperatures, interstellar reddening coefficient \(A_{V}\), distance \(D\), age \(t\), mass \(M/M_{\odot}\), equivalent width of the H\(\alpha\) line emission, H\(\alpha\) line width at 10\% intensity \(W_{10}\), and the corresponding literary source. Individual stellar parameters show discrepancies across different studies, which are likely due to the object's variability. For a detailed study of the cause of the variability of the spectrum and brightness of the star, it is necessary to conduct consistent observations over a long period of time.

In this work, we present the results of a new systematic long-term photometric and spectroscopic investigation of the star.

\begin{table}
\centering
\setlength{\tabcolsep}{1pt}
\small
\caption{Parameters of the star AS 442A, collected from the literature data.\label{tab1}}
\begin{tabular}{ccccccccc}
\hline\noalign{\smallskip}
Sp type & $T_{\text{eff}}$ (K) & $A_v$ & $D$ (pc) & $t$ (Myr) & $M/M{\odot}$ & $EW$ H$\alpha$ (\AA) & $W_{10}$ H$\alpha$ (km/s) & References \\
\hline\noalign{\smallskip}
B8Ve & 12500 & 3.85 & & & & & & \cite{Mathew2018} \\
& 11000 & & 826 & 1.5 & & & &\cite{Corporon1999} \\
& & & & & 3.5 & $-32.7$ & 646 [0.05] & \cite{Montes2009} \\
B8.5V & & & 700 & & & & & \cite{Pirzkal1997} \\
B8 & 10750 & 2.5 & 820.7 & 1.0 & 4.13 & & & \cite{Guzman-Diaz2021} \\
B8V & 11900 & & 834 & & & & & \cite{Thomas2023} \\
& 11000 & 2.263 & 859.8 & 0.84 & 3.89 & $-32.7$ & & \cite{Vioque2023} \\
B9/A0e & & & & & & & & \cite{Herbig1958} \\
B3e & & 1.74 & & & & $-18.5$ & & \cite{Cohen1979} \\
B8Ve & & & & & & $-23$ & & \cite{Finkenzeller1984} \\
\noalign{\smallskip}\hline
\end{tabular}
\end{table}

\section{Photometric studies}
Our requirements for the photometric data are as follows: a) a statistically significant number of measurements covering as long a period of time as possible; b) the data must be homogeneous and free of systematic errors. Observational data satisfying these conditions were taken from two archives. The first source contains photoelectric photometry data, while the second source provides CCD photometry results. These data sets are discussed separately below.

\subsection{Photoelectric photometry}
\label{sec21}
Photoelectric \(UBVR\) photometric observations of the star were performed at the Maidanak Observatory in Uzbekistan using the ROTOR program. The results of these observations are collected in the Strasbourg CDS archive \citep{herbst1994, grankin2007}. The photometric archive of young stars is available at \url{http://cdsweb.u-strasbg.fr/cgi-bin/qcat?J/A+A/461/183}. Observations of the HAeBe type star AS~442A were carried out regularly for 14 consecutive years during JD\,2445879--2450791 (1984--1997). In different years, between 60 and 150 measurements were taken per year. During this range of time, the number of measurements of the star AS~442A in the \(V\) and \(R\) bands was more than 1300, and slightly less in the \(B\) and \(U\) bands. Typical measurement errors are \(\pm \num{0.01}\) mag in the \(BVR\) bands and \(\pm \num{0.05}\)\,mag in the \(U\) band.

\begin{figure}[!htb]
   \centering
   \includegraphics[width=15.0cm, angle=0]{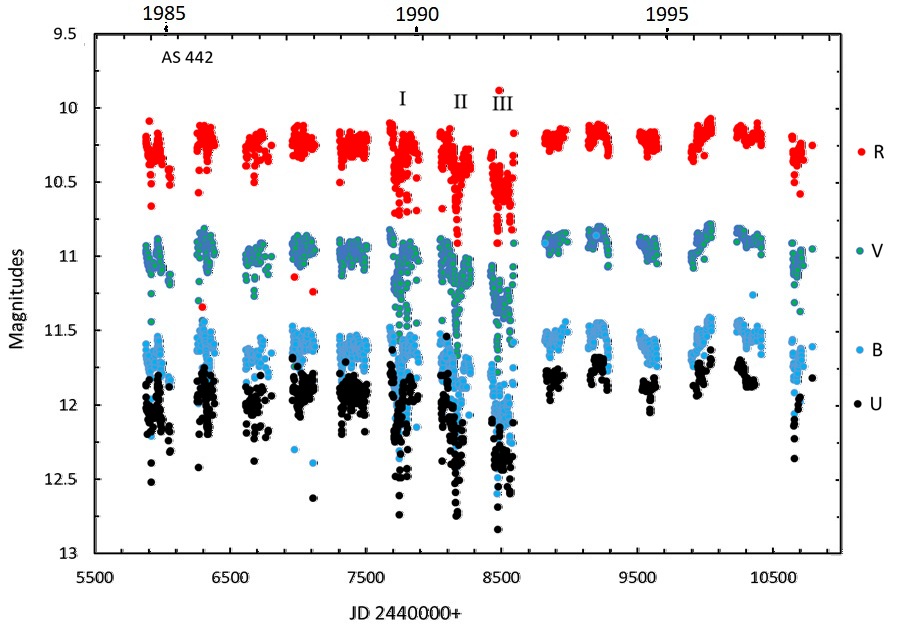}
\caption{14-year light curves in the \(UBVR\) bands of the star AS~442A. Different colors indicate stellar magnitudes in different photometric bands. The positions of three successive deep light drops are designated with I-III. }
   \label{Fig1}
   \end{figure}

\begin{figure}[!htb]
   \centering
   \includegraphics[width=14.0cm, angle=0]{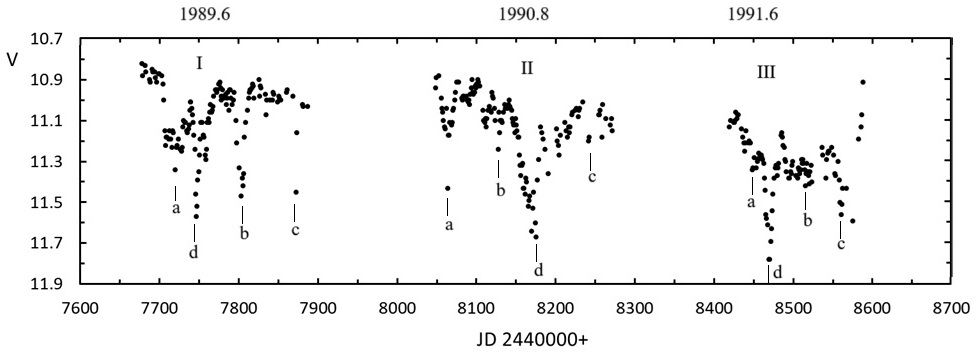}
\caption{Selected deepest drops in \(V\)-light. Dates of individual "components" of the dip are designated by the letters a, b, c, and d, where d is the date of the deepest drop. The Gregorian dates of the deepest drops are presented in the upper part of the diagram. }
   \label{Fig2}
   \end{figure}

Figure~\ref{Fig1} illustrates the star's \(UBVR\) light curves over the entire observation time, shown in different colors. In general, the star's brightness exhibits random variations around a roughly constant mean value in different seasons. Over the entire observation time of the star, only during 3 consecutive seasons, JD\,2447677 (I), JD\,2448166 (II), and JD\,2448588 (III) (1989--1991), does the star exhibit typical UX Ori (UXOR) drops (Fig.~\ref{Fig2}). The maximal amplitudes of the light drops are \(\Delta U \sim \num{1}\)\,mag, \(\Delta B \sim \num{0.8}\)\,mag, \(\Delta V \sim \num{0.7}\)\,mag, and \(\Delta R \sim \num{0.4}\)\,mag, respectively.
   
Since the nature of brightness changes in all bands is identical, subsequent analyses of periodicity focus exclusively on the \(V\)-band data.
Figure~\ref{Fig2} shows, on a larger scale, the I-III dips in the \(V\) magnitude of the star observed successively in 1989--1991. As can be seen, in addition to the deepest minimum for each year, designated by the symbol d, relatively shallow brightness drops are also evident, with dates designated by the letters a, b, and c. The Gregorian dates of the deepest drops are presented in the upper part of Figure~\ref{Fig2}.
The time intervals between the I and II drops are approximately \num{429} days, and between the II and III drops \(\sim \num{295}\) days in all \(UBVR\) bands.

As shown in Figure~\ref{Fig2}, during the appearance of the three main deep drops, four additional minima of smaller amplitude were identified in each case, located near the principal brightness decline. From this, it is clear that the time intervals between the dips do not exhibit any repeatable pattern. Typical observed intervals between dips range from \num{22} to \num{69} days.

\begin{figure}
   \centering
   \includegraphics[width=10.0cm, angle=0]{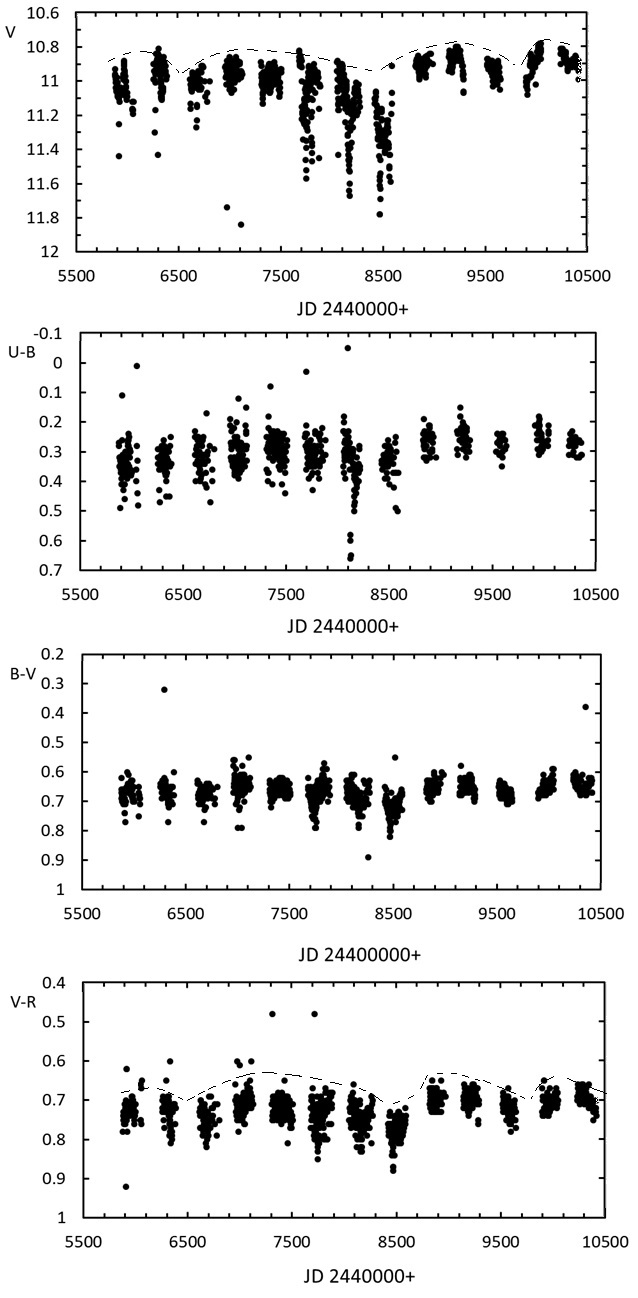}
\caption{From top to bottom—the 14-year \(V\) light curve and \(U-B\), \(B-V\), and \(V-R\) color indices variations over time. Each cluster of points corresponds to data from a single year (season) of observations. The dashed lines characterize the bright state of the star. }
   \label{Fig3}
   \end{figure}

Figure~\ref{Fig3} shows the graphs of time variations of the colors \(U-B\), \(B-V\), and \(V-R\) and \(V\) magnitudes. As can be seen, with practically constant average color indices (\(\langle U-B \rangle =\num{0.3}\), \(\langle B-V \rangle = \num{0.55}\), \(\langle V-R \rangle =\num{0.74}\)) in different seasons, their seasonal fluctuations exhibit an amplitude of approximately \num{0.2}\,mag in all colors near the average level. Note that the nature of the variation in color indices is preserved in all seasons, even in time, when the mentioned above drops (I-III) are observed. In Figure~\ref{Fig3}, the dashed lines represent the bright state of the star. In the light curve and colors \(U-B\), \(V-R\), and weakly in \(B-V\), smooth wave-like seasonal variations in the average magnitude and colors with amplitudes of about \num{0.2}\,mag are observed.

\indent In the light curve and in the \(U-B\), \(V-R\), and, to a lesser extent, \(B-V\) indices, smooth, wave-like seasonal variations in the mean stellar magnitude and colors with amplitudes of about \num{0.2}\, mag are observed.

\indent As shown, the smooth long-term variation in \(V\) brightness and colors during JD\,2446682--2448471 spans approximately six years, while the interval JD\,2448471--2449644 covers about four years.

\begin{figure}
\centering
\includegraphics[width=15cm, angle=0]{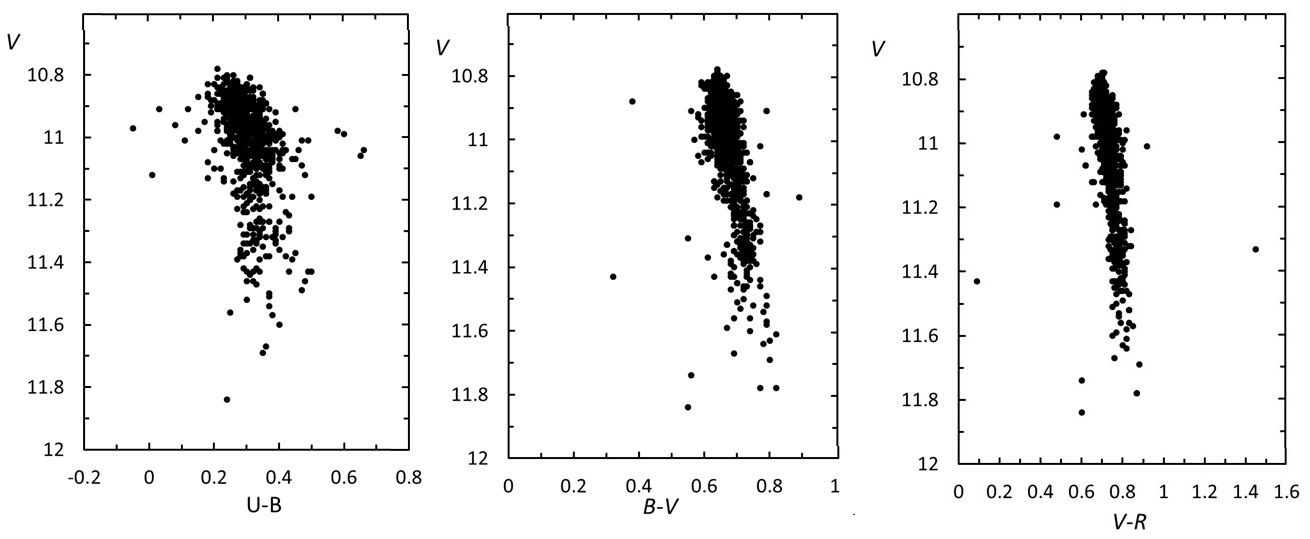}
\caption{Color-magnitude diagrams of \(U-B\), \(B-V\) and \(V-R\) versus \(V\) magnitude of the star AS~442A.}
   \label{Fig4}
   \end{figure}

Figure~\ref{Fig4} presents color–magnitude diagrams of \(U-B\), \(B-V\), and \(V-R\) versus \(V\) magnitude. The star generally becomes redder as it fades, suggesting normal interstellar extinction. When the star's brightness is greater than \(V \sim \num{11.1}\)\, mag, all colors demonstrate nearly linear dependence on brightness. However, when the brightness lies in the range \num{11.1}--\num{11.8}\, mag, the \(U-B\) index and, more weakly, the other colors lose their dependence on brightness. At this stage, the ``blueing effect'' is observed, a phenomenon well known in UXOR stars \citep{Grinin1994, herbst1999, semkov2015, huang2019}.

According to the dust-grain occultation model, the observed color variation is caused by scattered light from small dust particles. Typically, a star becomes redder when its light is blocked by dust grains along the line of sight. However, when the degree of occultation increases sufficiently, the fraction of scattered light in the total observed flux becomes significant, causing the star to appear bluer.

\subsection{CCD photometry}

\begin{figure}
\centering
\includegraphics[width=14cm, angle=0]{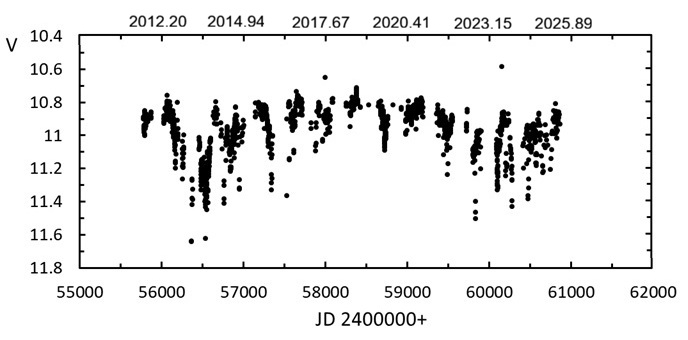}
\caption{\(V\)-light curve of the star AS~442A based on the AAVSO data in the time interval 2011-2025}
   \label{Fig5}
   \end{figure}

The second part of the data was downloaded from the archive of the American Association of Variable Star Observers (AAVSO, \url{https://www.aavso.org/}). Although this collection contains over 2200 CCD \(V\)-magnitude values, the data for other bands are much smaller and often obtained at different times. The typical \(V\)-magnitude measurement error is \(\pm \num{0.003}\)\,mag. Therefore, from the collected data, we used only the \(V\)-magnitude values, excluding the visual magnitude data and a small amount of data in other photometric bands. Figure~\ref{Fig5} shows the light curve of AS~442A based on AAVSO data, covering the period from 2011 to 2025 (a 14-year observational span).

As shown, two deep dips in brightness are observed, each lasting approximately 5--6 years, along with other less significant dips. Interestingly, each of the two deep dips, as in the first case (see Section~\ref{sec21}), is accompanied by several dips occurring before and after the deepest minimum. The time interval between the closest dips in brightness is about \num{80} days. Since observations were continuous (Fig.~\ref{Fig5}), the second deep drop in light occurred approximately \num{3568} days after the first deepest minimum.

\section{Analysis of periodicity in brightness}

To detect the periodic component in the brightness variations of AS~442A, the first stage of the analysis focused on searching for periods shorter than one year. For this purpose, we divided the overall light curve into separate arrays. Each array consists of data obtained during a single observing season. The main rationale for this approach is that it is more appropriate to use points from a single season rather than the entire 14-year dataset. This strategy minimizes the distortion introduced by stochastic components in the brightness variations.
For statistical Fourier analysis, we used the Period 04 program \citep{lenz2005}, which was developed based on the methods of \citet{lomb1976} and \citet{scargle1982}. The program applies Fourier analysis to identify the most probable periods in the power spectrum.

\begin{figure}
\centering
\includegraphics[width=14cm, angle=0]{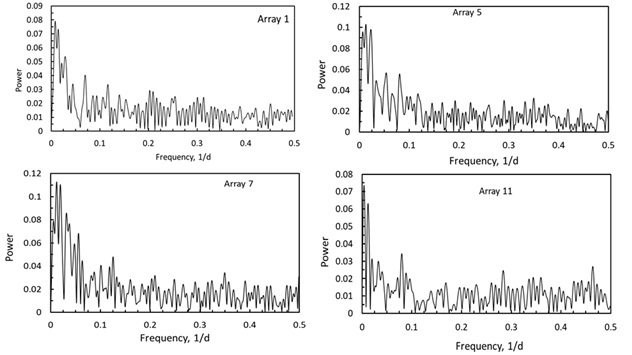}
\caption{ Examples of power spectra obtained for different arrays.}
   \label{Fig6}
   \end{figure}

Data obtained in individual observation seasons do not allow us to reliably identify periods that are clearly and consistently distinguished in the power spectrum. To identify periods, we considered those peaks whose powers in the spectrum exceeded, as a minimum, the \(3\sigma\) level, where \(\sigma\) is the root mean square (rms) of the "noise" power signal. In addition, we checked whether the most probable frequency was also present in the spectra of other arrays. As an example, Figure~\ref{Fig6} shows the power spectra obtained for different arrays. We did not detect the previously reported \num{16.2} day period with an amplitude of \num{0.2}\,mag, as noted by \citet{melnikov1993}.

\begin{figure}
\centering
\includegraphics[width=12cm, angle=0]{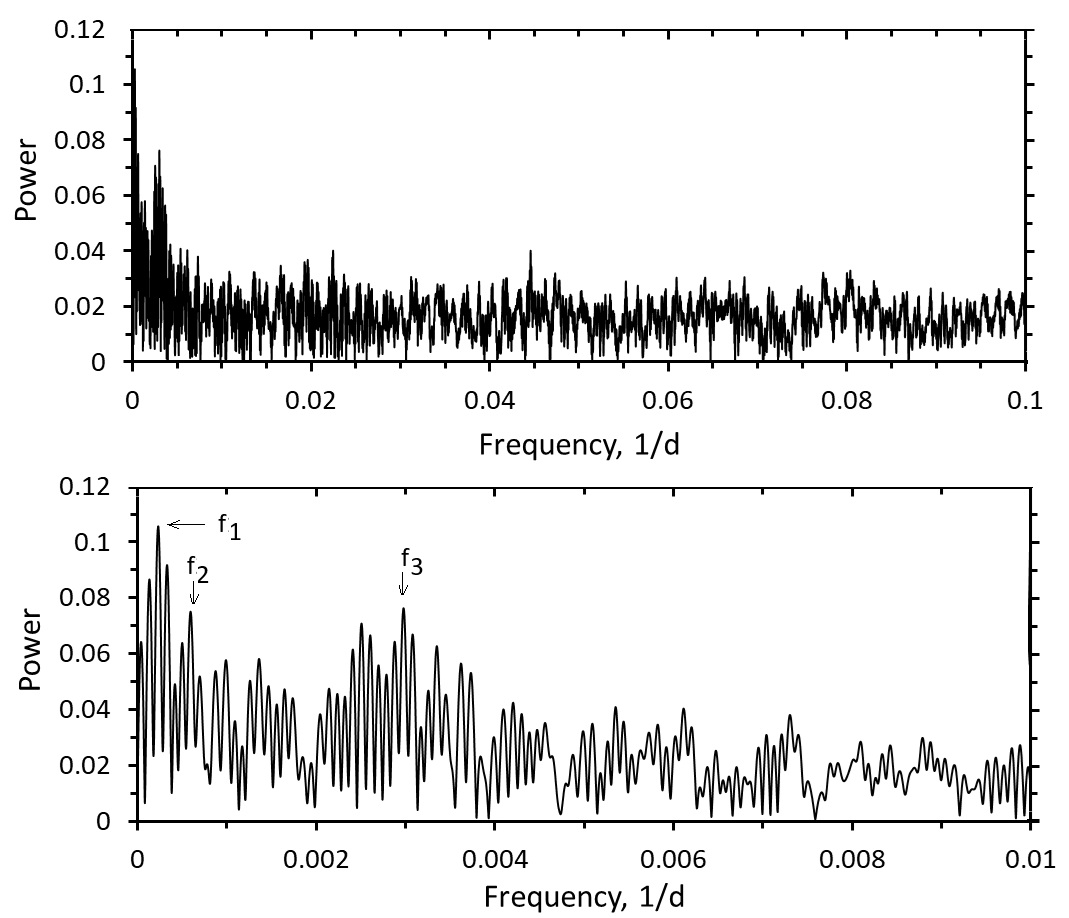}
\caption{Power spectrum based on all \(V\)-magnitude data for the frequency intervals 0-0.1\,d\(^{-1}\) (top panel) and 0-0.01\,d\(^{-1}\) (bottom panel). The arrows mark significant peaks corresponding to \(f_1\), \(f_2\), and \(f_3\) (see text for values).}
   \label{Fig7}
   \end{figure}

The study showed that several periods were repeatedly detected in 3--5 individual arrays. Of these, periods of \(P_1 = 50 \pm 3\) days (in 5 arrays), \(P_2 = 76 \pm 2\) days (in 4 arrays), and \(P_3 = 80 \pm 2\) days (in 2 arrays) were obtained. Figure~\ref{Fig7} shows that the probable periods identified in most datasets fall primarily within the range of 50--80 days. However, these periods, although evident in some arrays, are not consistently observed across other years.

As seen from Figure~\ref{Fig3} and Figure~\ref{Fig5}, in the bright state, a smooth long-term variability in the star's brightness is observed. Therefore, it is necessary to check the existence of a long-term periodic component in the star's brightness. Figure~\ref{Fig7} shows an example of the power spectrum constructed based on all available \(V\)-magnitude data. The upper panel shows the power spectrum in the frequency range 0--0.1 d\(^{-1}\), and the second panel in the range 0--0.01 d\(^{-1}\). As can be seen, in the low-frequency region there are several significant peaks, \(f_1 = 0.000266 \pm 0.000002\) d\(^{-1}\), \(f_2 = 0.000334 \pm 0.000001\) d\(^{-1}\), \(f_3 = 0.002980 \pm 0.000003\) d\(^{-1}\), marked with arrows. These peaks correspond to the periods \(P_1 = 3759 \pm 35\) days, \(P_2 = 2995 \pm 9\) days, and \(P_3 = 335.57 \pm 0.68\) days, respectively.

The most probable of them is the period \(P_1\), since this value of the period \(P_1 = 3759 \pm 35\) days is equal to approximately \(10.3 \pm 0.1\) years. Note that the time interval between the deepest drop, which was observed in 1990 (Fig.~\ref{Fig3}), and the subsequent drops obtained in 2012.5 (21.5 years) and 2023 (32 years) (Fig.~\ref{Fig5}) is a multiple of this period. Thus, we report for the first time that all three deep minima occurred at intervals consistent with \(P_1\). This shows that the observed deep drops in the star's brightness recur with a 10.3-year cycle.

\begin{figure}
\centering
\includegraphics[width=14cm, angle=0]{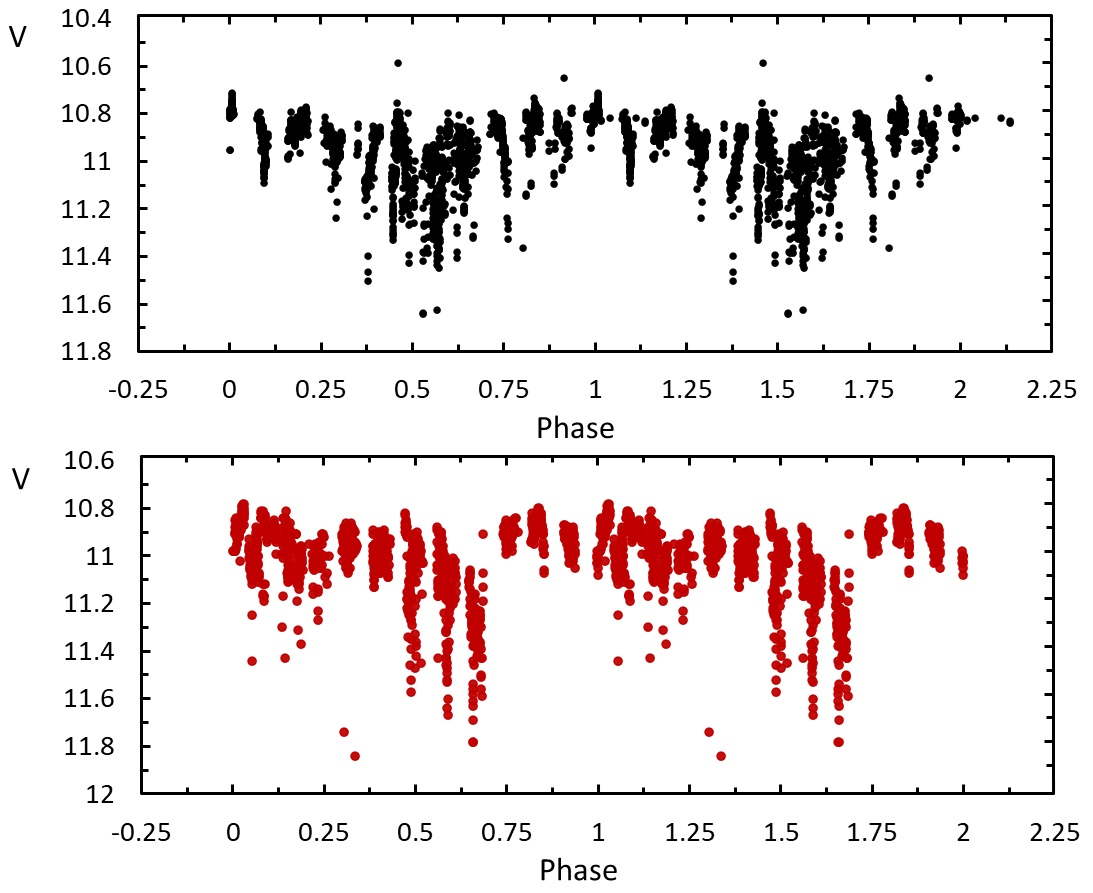}
\caption{Phased V-light curves on the data of AAVSO (top panel) and on the data of ROTOR (bottom panel).}
   \label{Fig8}
   \end{figure}

In Figure~\ref{Fig8}, the phased \(V\)-light curve of the star for the period of 10.3 years is illustrated, with elements \(V_{\text{max}} = \text{JD}\,2458352.53 + 3759E\). The upper panel shows the AAVSO data, and the lower panel shows the ROTOR data. As can be seen, in the AAVSO data there are two deep minima (Fig.~\ref{Fig5}), and in the ROTOR data there is one deep minimum (Fig.~\ref{Fig3}), which coincide closely in phase with this period.

\section{Spectral observations}

Spectral observations of AS~442A were conducted at the Cassegrain focus of the Shamakhy Astrophysical Observatory (ShAO) 2-m telescope using the Modified Universal Astro Grid Spectrograph (MUAGS). The detector was an Andor CCD (iKonL-936-BEX2-DD) with 2048$\times$2048 pixels (13.5\,$\mu$m pixel size), coupled to the spectrograph via a Canon EF fast lens (f=200 mm, f/2). Focal length of the collimator is 1100 mm, camera's focal length was 200 mm, angle of incidence $\alpha$ and diffraction angle $\beta$ are 27.5$^\circ$ and 20.5$^\circ$, respectively. The width of two pixels determines the resolution in the camera focal plane, so for binning 1$\times$1, the monochromatic image of the slit is S$'$ = 2 px = 0.027 mm. A diffraction grating of 651 line/mm was used, which in the first order and at binning 1$\times$1 gives the spectrum with a linear inverse dispersion of 144\,\AA/mm in the range of $\lambda$ 3600--8000\,\AA. In other words, in the region of the H$\alpha$ line ($\lambda$ 6562.816\,\AA) in this mode, we have obtained a moderate spectral resolution of about R=3400. A more detailed description of the spectrograph and the observation method is given in \citet{ismailov2023}.

The entire process of observation and data reduction was performed using the DECH program \url{http://www.gazinur.com/DECH-software.html}. Table~\ref{table2} presents the average UT time for the acquisition of two spectrograms per star. After calibration and processing, all spectra were averaged and normalized to the continuum. 

Our spectral observations were executed for 22 nights from 2020 to 2024. Table~\ref{table2} presents the observation log of the star AS~442A. The columns, from left to right, list the observation date, Julian date, exposure time, binning, and signal-to-noise relation S/N at the H$\alpha$ line region in the spectrum. After calibration and processing, all spectra were averaged for the nights and normalized to the continuum. For the spectra of standard stars, the standard deviation in measurements for the radial velocities (RV) of strong lines is $\pm$ 5 km/s, and for the equivalent widths ($EW$) of hydrogen lines for A0--A6 type stars, is $\pm$ 0.5\,\AA.

\begin{table}
\centering
\setlength{\tabcolsep}{5pt}
\small
\caption{The log of spectral observations of AS 442A.\label{table2}}
\begin{tabular}{lcccc}
\hline\noalign{\smallskip}
Date & JD 2450000+ & Exp (s) & Bin & S/N \\
\hline\noalign{\smallskip}
16.07.2020 & 9047.446 & 300 & 3x3 & 80 \\
19.07.2020 & 9050.455 & 700 & 2x2 & 95 \\
24.07.2020 & 9055.479 & 700 & 2x2 & 87 \\
19.08.2020 & 9080.339 & 700 & 2x2 & 85 \\
05.08.2021 & 9432.341 & 700 & 2x2 & 84 \\
06.08.2021 & 9433.276 & 700 & 2x2 & 89 \\
10.09.2021 & 9468.331 & 700 & 2x2 & 77 \\
13.09.2021 & 9471.298 & 700 & 2x2 & 68 \\
21.06.2022 & 9752.406 & 1200 & 1x1 & 127 \\
30.06.2022 & 9761.345 & 1200 & 1x1 & 113 \\
01.07.2022 & 9762.383 & 1200 & 1x1 & 128 \\
25.08.2022 & 9817.276 & 2400 & 1x1 & 133 \\
26.08.2022 & 9818.255 & 2000 & 1x1 & 120 \\
27.08.2022 & 9819.253 & 1800 & 1x1 & 121 \\
28.08.2022 & 9820.24 & 1800 & 1x1 & 114 \\
29.08.2022 & 9821.238 & 1800 & 1x1 & 116 \\
30.08.2022 & 9822.242 & 1800 & 1x1 & 115 \\
31.08.2022 & 9823.251 & 1800 & 1x1 & 110 \\
16.09.2022 & 9839.3104 & 1800 & 1x1 & 120 \\
02.10.2022 & 9855.225 & 2400 & 1x1 & 118 \\
04.10.2022 & 9857.244 & 2400 & 1x1 & 120 \\
08.09.2024 & 10562.214 & 1500 & 1x1 & 115 \\
\noalign{\smallskip}\hline
\end{tabular}
\end{table}

\begin{figure}[htbp]
\centering
\includegraphics[width=15cm, angle=0]{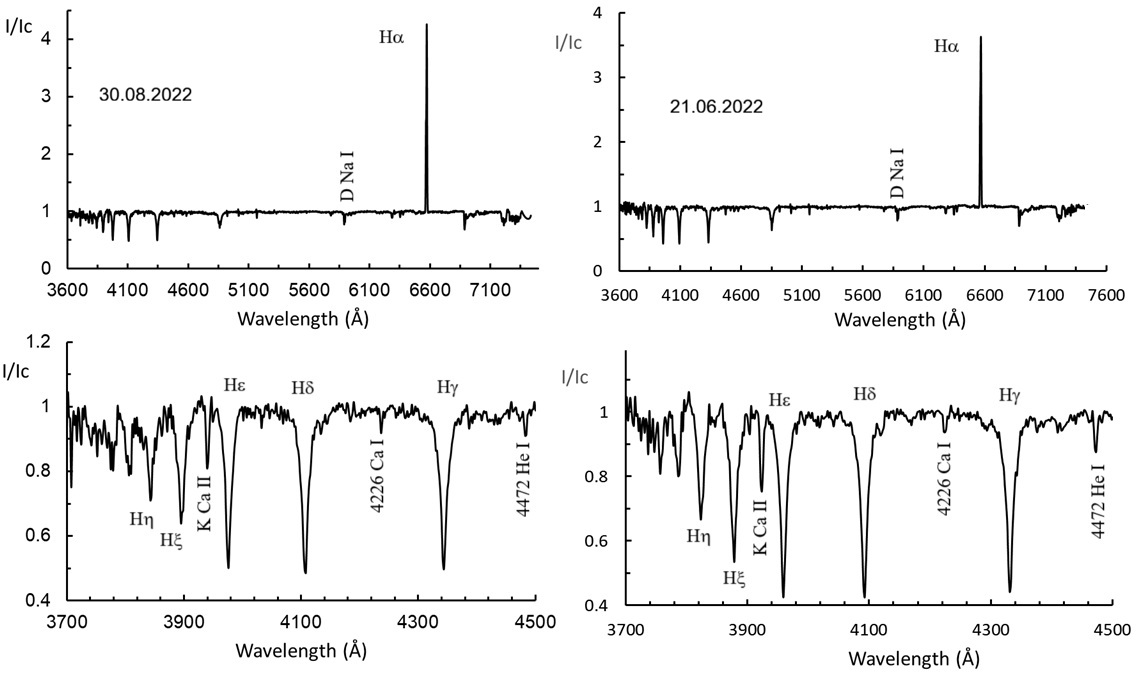}
\caption{An overview of the spectrum of the star AS 442A is shown for the dates 21.06.2022 and 30.08.2022. The following panels show a larger-scale representation of the spectrum on those dates in the range $\lambda$\,3700-4500\,\AA.}
   \label{Fig9}
   \end{figure}

Figure~\ref{Fig9} shows an example of the star's spectra obtained on 21.06.2022 and 30.08.2022, as well as a fragment of these spectra in the range $\lambda$ 3700--4500\,\AA\, on a larger scale. As can be seen, in addition to the strong Balmer series lines, the spectrum also contains Na\,I D, He\,I, Ca\,I, Ca\,II K, Fe\,II 4924, and other lines. The H$\alpha$ line is a strong emission, and a weak emission component is also observed at the center of the H$\beta$ line. No traces of emission were detected in the other hydrogen lines of the Balmer series.

The difference in the absorption spectrum on the two dates, shown on the same scale, is clearly visible: on the night of 21.06.2022, the absorption lines were significantly stronger.

\begin{figure}
\centering
\includegraphics[width=14cm, angle=0]{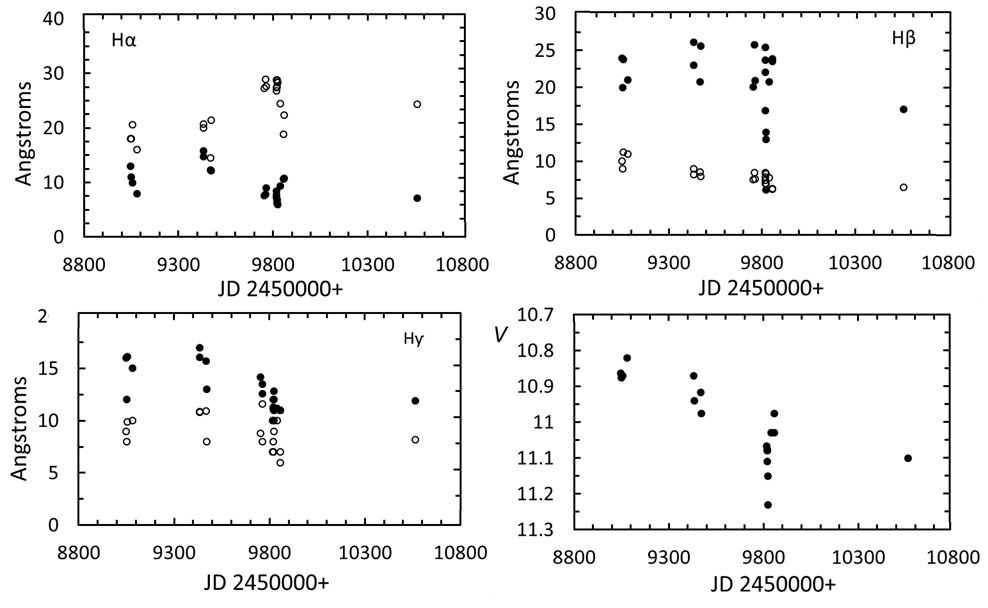}
\caption{Time variations of the equivalent widths (open circles) and $FWHM$ (black circles) of the H$\alpha$-H$\gamma$ lines. The H$\alpha$ is the emission line; all remaining lines are in absorption. In the last right panel, the time variation of the $V$-band magnitude is presented.}
   \label{Fig10}
   \end{figure}

Figure~\ref{Fig10} illustrates the time variations of the equivalent widths (open circles) and full widths at half maximum ($FWHM$) (filled circles) of the hydrogen lines H$\alpha$, H$\beta$, and H$\gamma$ over our spectral observation period. One of the panels of Figure~\ref{Fig10} shows $V$-magnitude variations for the same time as our spectral observations. From this, it is evident that the radiation of the star over the entire season of spectral observations shows rapid daily fluctuations, and the average brightness decreases in the range of $\Delta V = 0.5$ mag. According to Figure~\ref{Fig5}, the last deep dip in brightness of the star occurs around 2023. Our spectral observations were carried out at the beginning of the brightness weakening, when, especially in 2022, the $V$-magnitude of the star decreased by $\sim$0.5 mag.

As can be seen from Figure~\ref{Fig10}, the annual average equivalent width of the H$\alpha$ emission line varies from $-$15 to $-$30\,\AA\ \text{and} increases smoothly from 2020 to 2022. In addition to the long-term variations, the 2022 data clearly show that the equivalent widths of the H$\alpha$ emission line and $FWHM$ also exhibit relatively rapid changes in time scale of several days, which is a typical property of HAeBe type stars.

Figure~\ref{Fig10} shows that the $EW$ of H$\alpha$ emission is increasing and simultaneously accompanied by a decrease in the $FWHM$. Concurrently, the strengthening of the emission is associated with a decrease in both the $EW$s and $FWHM$s of the H$\beta$--H$\gamma$ absorption lines. At the same time, such variations in the spectrum are accompanied by a synchronous weakening in the brightness of the star.

We considered the dependence of the spectrophotometric parameters of individual spectral lines versus the stellar magnitude.
Appendix Figure~\ref{Fig1A} shows the relationship between the $EW$ and $FWHM$ of the H$\alpha$, H$\beta$, H$\delta$, and Fe II 5018 spectral lines versus the $V$-magnitude. These diagrams clearly show that, with increasing $V$-magnitudes, the $EW$ increases and the $FWHM$ of the H$\alpha$ emission line decreases. With increasing $V$-magnitudes, all other hydrogen absorption lines H$\beta$--H$\delta$, as well as the Fe II 5018 absorption lines, show little minor fluctuations with a constant mean value of the $EW$. At the same time, the $FWHM$ of all these lines showed an increase when the star's $V$-magnitude decreased.

In Appendix Figure~\ref{Fig2A}, the relation of the line depth $R$ of spectral lines H$\alpha$--H$\delta$ and Fe II 4924, 5018, Mg I 5172 versus $V$ magnitude is presented. Here, it is clear that with an increase in stellar brightness, all absorption lines show a decrease in line depth, approximately twice.

We denote the flux of radiation in the $V$ band at maximum magnitude as $F_{1}$, and at minimum magnitude as $F_{2}$. Then, for a change in brightness by a value of $\Delta m$ of about 0.5 mag, we obtain $F_{1}/F_{2} = 0.63$. This means that at a bright state of brightness, the flux of radiation in the continuum is approximately 1.6 times greater than at a weak condition.

The equivalent width of the H$\alpha$ emission line is defined relative to the continuum \(I_{C}\) and line \(I\) intensities as:

\begin{equation}
EW = - \int_{\lambda_{1}}^{\lambda_{2}} \frac{I - I_{c}}{I_{c}} \, \mathrm{d}\lambda
\label{eq1}
\end{equation}
\noindent where $\lambda_{1}$ and $\lambda_{2}$ are the boundaries of the range covered by the line at the continuum level.
As the stellar brightness increases (reflecting a rise in the continuum level), the denominator of this expression grows. Assuming the intrinsic line flux remains constant, the measured value will decrease accordingly. Our data, particularly the observed correlation with the photometric state, suggest that continuum variability is the primary driver of the fluctuations. Furthermore, the simultaneous increase in $FWHM$ indicates that high-velocity processes, likely originating from the innermost disk regions or the base of the stellar wind, become more prominent during the bright state.

The increase in $FWHM$ indicates that at moments of increased brightness, a broadening of the spectral lines and a decrease in their residual intensities occurred. This may be associated with the occurrence of high velocities in the inner parts of the disk or base of the wind.

Figure~\ref{Fig3A} in the Appendix presents selected H$\beta$ line profiles obtained during the 2022 observing season. Significant short-term profile variations are evident on timescales of several days. Similar to H$\alpha$, the H$\beta$ profiles exhibit two weak emission peaks. Between JD\,2459752 (June 21, 2022) and JD\,2459818 (August 26, 2022), these weak emission components are superimposed on a broad absorption line. Subsequently, from JD\,2459823 (August 31, 2022) to JD\,2459857 (October 4, 2022), the line transitions into a typical P Cyg profile, characterized by deep blueshifted absorption and a strong emission component. These spectral transformations support our assumption: towards the end of the 2022 season, the significant decrease in stellar brightness is accompanied by an increase in H$\beta$ emission intensity and a concomitant reduction in the $FWHM$.

\section{Discussion}
In this paper, we perform a multi-year photometric and spectral study of the Herbig AeBe star AS~442A with a complex circumstellar structure. The star is a physical pair with the secondary component AS~442B, with a brightness of $V \sim 15.8$ mag and a spectral type of G5/K0e \citep{Cohen1979}, located at a distance of 4.95 arcseconds from the primary \citep{Thomas2023}. AS~442A has a brightness of about $V \sim 11.5$ mag, so the contribution of the secondary component to the brightness and spectrum of the primary is insignificant ($\sim 1.9\%$). Therefore, the observed photometric and spectral features are mainly attributed to AS~442A.
 
The analysis of the star's light curve was performed using more than 1300 data points obtained over 14 years of observations, and additionally, AAVSO data comprising more than 2000 points obtained over 15 years. It is shown that the star's brightness primarily exhibits the characteristics typical of young UXOR stars. In addition to the random brightness variability typical of UXOR-type stars, we found a long-time periodic variability with an amplitude of \(\Delta V \sim \num{1}\) mag.

In its bright state, AS~442A exhibits rapid light fluctuations with amplitudes no greater than a few tenths of a stellar magnitude and characteristic time scales of one to several days (Figure~\ref{Fig1}). 
Despite such rapid variability, no stable short-term periodic components were detected within individual observing seasons. Statistical periodogram analysis revealed that the star exhibits long-term periodic brightness variability with a period of $P = 10.3 \pm 0.1$ yr. This appears to be a unique feature for a UXOR-type star.

Unfortunately, the spectral resolution of our spectra does not allow us to estimate the precise values of the radial velocities of the spectral lines. In the work \citet{Corporon1999} for AS~442A, based on insufficiently reliable data, a possible spectroscopic period of 64 days was determined. 

The 10.3-year period revealed in this work on the photometric deep drops for the star AS~442A represents a unique new result among UXOR-type Herbig Ae/Be stars. A similar 10-year period was first detected for the classical UXOR type T Tauri star CQ Tau \citep{Grinin2023}. Our data confirm the existence of multi-year periodic brightness variations in UXOR stars, which were previously thought to be sporadic.

As shown in Figure~\ref{Fig2}, each deep dip in brightness is accompanied by smaller dips with unstable time intervals of approximately 22--69 days (Section~2.1). We suggest that the observed timescales of the brightness declines are well explained by the passage of stable gas-dust clumps within the stellar disk across the line of sight. Before the main deep eclipse, which is likely caused by a total or partial eclipse of the primary star by the companion, the primary star is also sometimes eclipsed by these dense dark clumps located on Keplerian orbits within the primary star's circumstellar disk.

We apply Kepler's third law for known values of the long-term period of 10.3 years and the mass of the primary star, which is $\sim 3.5\, M_{\odot}$ \citep{Montes2009}, and if we assume that the reasonable mass of the secondary component is in the range of $\sim 0.5$--$4\, M_{\odot}$, and using the following known formula, we can obtain the possible semi-major axis of the orbit of the secondary component:

\begin{equation}
a = \sqrt[3]{P^2 (M_a + M_b)}
\label{eq2}
\end{equation}

\noindent Here $a$ is the semi-major axis in astronomical units (AU), $P$ is the period in years, $M_a$ and $M_b$ are the masses of the primary and secondary components, expressed in solar mass $M_{\odot}$. From these results, we derive the semi-major axis $A$ ranging from $7.51 \pm 0.1$ to $9.3$ AU. This suggests that the long-term variability characterized by brightness dips may be associated with the presence of a close companion located at a minimum distance of $7.51$ AU.

From Eq.~\ref{eq2}, it is clear that even if the secondary component has a larger mass, the semi-major axis of the binary system cannot be larger than $10$ AU. However, what can be the semi-major axis of the known AS~442B component?

\begin{equation}
a(\text{AS 442B}) = d(\text{arcsec}) \times D(\text{pc})
\label{eq3}
\end{equation}

By using the relation for the visual binary system, the distance between the components $d = 4.75''$ and the known distance to the target $D = 834$ pc (\citet{Thomas2023}), we can obtain the semi-major axis corresponding to the AS 442B component as $3961$ AU. So, this distance does not correspond to the observed $10.3$-year period. This analysis shows that the previously discovered AS 442B component around the star cannot cause a $10.3$-year periodic smooth change in the brightness of the star. This is most likely due to the presence of a smaller, low mass star with a semi-major axis of $A \geq 7.51$ AU or a few big mass planets.

The measured values of radiation polarization for AS 442A were presented in two papers. In the paper by \cite{Oudmaijer2001}, based on three estimates obtained in 1998, P=2.9\%, corresponding to  phases 0.03-0.05 (for light curve phase elements see section 3); on the phase light curve, this corresponds to a bright state. In the paper by \cite{Petrova1987}, polarimetric observations were carried out on 14.08.1985 (JD 2446291.522), which corresponds to phases 0.72, i.e., again to a bright state of brightness (see Figure \ref{Fig8}). In the work \cite{Petrova1987}, the polarization in the $UBVRI$ bands varies smoothly from 2.67\% to 2.31\% with a prominent maximum of 2.98\% in the V band. These values are in good agreement with the data of \cite{Oudmaijer2001} within the error limits. Thus, during the bright state, polarization is consistently observed within the range of 2.5-3.0\%. During moments of deep dips in the star's brightness, polarization was not measured.

Indeed, the polarization degree of $2-3$\% observed in AS 442 is quite high for HAeBe stars. This value provides direct evidence for the presence of a non-spherical, asymmetric dust disk surrounding the star. Typically, such high polarization in HAeBe stars arises from the following factors:
\begin{itemize}
\item[a)] If the disk is viewed at a significant inclination (near edge-on), strong polarization is produced by the scattering of stellar light off dust grains within the disk;
\item[b)] A polarization degree of 2--3\% significantly exceeds the values typically attributed to the interstellar medium, implying that the primary source of this effect is the circumstellar environment;
\item[c)] The fact that the polarization remains as high as 2--3\% even in the bright state suggests that the disk is oriented at a relatively large inclination angle to our line of sight.
\end{itemize}

Our results indicate that the sudden brightness drop in UXOR stars may not be random, but rather the result of the presence of a secondary companion. So far, such a pattern has only been observed in two objects: CQ Tau \citep{Grinin2023} and AS 442A (in our study). This suggests conducting studies on the periodicity of brightness variations in other UXOR-type stars. This requires multi-epoch observations spanning a protracted time baseline.

The interferometric measurements by \cite{Eisner2004} showed that perhaps the resolved circumstellar material around AS 442A may be significantly inclined. IR observations at 2.2 $\mu$m in this work provide an inclination of the disk of nearly \ang{46}--\ang{48}. On the data of \cite{Eisner2004} for CQ Tau, the inclination angle of the inner region of the CS disk to the line of sight was equal to \ang{48}. Moreover, in the sub mm region, the CS disk was observed almost pole-on \citep{Chapillon2008}. Recently, this result for CQ Tau was confirmed in ALMA interferometer observations \citep{Ubeira2019}. These data indicate that, despite the relatively high inclination of the disk, both CQ Tau and AS 442A exhibit periodic eclipses of the central star with a period of approximately 10 years. The relatively small amplitude of the dip observed in AS 442A can be explained by a slight tilt of the circumstellar disk relative to the observer's line of sight. In this case, the secondary component must be large enough to cause such an eclipse. Candidate objects for such a configuration include young T Tauri stars located within their own extended disks of gas and dust.

Photometric color-magnitude diagrams of the star showed that in the faint state, the star exhibits the ``blueing'' effect, which is the sign of UXOR stars \citep{Tambovtseva2025}. The first interpretation of the ``blueing'' event is based on the assumption that the event occurs in a binary system with a faint blue secondary companion \citep{Wenzel1969}. According to this hypothesis, when the star is eclipsed by dust clouds located in the circumstellar disk of the star, the radiation of the blue star becomes dominant. The alternative explanation of the ``blueing'' effect is based on the assumption that the source of the blue radiation of UXOR is the diffuse radiation of protoplanetary disks, the contribution of which increases during deep minima \citep{Grinin2023}. The presence of a close companion may naturally reconcile both interpretations, as a low-mass secondary located at several AU could potentially contribute additional blue flux while simultaneously perturbing the inner disk structure, thereby enhancing scattered light during deep minima.

It is important to distinguish between the short-term UXOR-type variability, which is most likely caused by stochastic dust occultation in an inclined inner disk, and the smooth, decade-scale 10.3-year modulation discussed above. While the former can be explained within the classical UXOR framework, the latter suggests the presence of a longer-term dynamical driver, such as a close companion perturbing the disk structure. In this context, the companion scenario may provide a unifying explanation for both the periodic modulation and the enhanced dust activity.

Since the $A_V$ parameter differs significantly according to different authors (see Table~\ref{tab1}), we decided to redefine its value based on a relatively more accurately determined spectral class and the observed $B-V$ color index of the star. Since the brightness and color indices of the star are changing in time, it would be logical to use the value corresponding to the maximum brightness state as the $B-V$ color index of the star \citep{Tambovtseva2025}. For a normal star of spectral class $B8V$, the color index value $(B-V)_0 = -0.109$ \citep{Pecaut2013}, with a bright state of the star $B-V = 0.64$, then the color excess $E(B-V) = B-V - (B-V)_0 = 0.75 \pm 0.01$. Taking the extinction coefficient values $R = 3.1$, we obtain $A_V = 2.32 \pm 0.03$. At a bright state of the star's brightness, $V_{\text{max}} = 10.79$, taking the distance to the object as 834 pc (Table~\ref{tab1}), bolometric correction coefficient $m_b$, and using the expression
\begin{equation}
    \begin{aligned}
        M_{vb} = V_{\text{max}} + 5 - 5 \log D - A_v + m_b
    \end{aligned} \label{eq3}
\end{equation}

\noindent we obtain $M_{vb} = -1.93 \pm 0.05$. Further, taking solar parameters $M_{vb\odot} = 4.7$, $T_{\odot} = 5800$ K, using for the star $T_* = 12500$ K \citep{Pecaut2013}, for the luminosity we have obtained $L_* = 448.7 \pm 0.7\,L_{\odot}$, and the radius of the star as $R = 4.56 \pm 0.5\,R_{\odot}$. The obtained parameters are in very good agreement with the data given in the catalog of \citet{Guzman-Diaz2021}.

A certain part of the photometric observations has been carried out (AAVSO data) synchronously with our spectral observations. This allowed us to track the spectral variability as a function of the star's brightness. Between 2020 and 2024, the maximal range of brightness variation in the $V$-band was approximately 0.5 mag. It was shown that during this $\sim 0.5$ mag decrease in brightness, the $EW$ of the absorption spectral lines remains unchanged, but the $FWHM$ gradually decreased, and the residual intensities $R$ increased. This indicates that the primary cause of the spectral line variations is a change in the continuum level due to brightness variations.

As the star's brightness increases, the equivalent width and depth of the emission line H$\alpha$ decrease. The $FWHM$ increases in the inferred line. As shown in Section~4, the reduction in the continuum level due to the fading brightness contributes to an increase in depth and a decrease in the width, which is observed in the star's spectrum. At the end of the 2022 spectral observing season, the stellar brightness decreased. During this period, an increase in emissions within the H$\alpha$ and H$\beta$ line profiles, accompanied by a more pronounced development of P Cyg profiles, indicated an intensification of the disk wind along the line of sight. These phenomena served as a precursor to the subsequent deep fading event observed in 2023.

\section{Conclusion}
Thus, the results of long-term photometric and spectral observations of the HAeBe star AS 442A allowed to identify some interesting characteristics of the star:
\begin{itemize}
    \item AS~442A was identified as a young Herbig Ae/Be star with properties of UXOR stars. Three distinct deep photometric minima with amplitudes of approximately $\Delta V \sim 1$ mag were detected during the period 1984--2025. The corresponding colour-magnitude diagrams demonstrate the typical ``blueing'' effect during these eclipses.
    
    \item Each of the three deep brightness minima detected is accompanied by relatively smaller brightness drops observed before and after the main deep minima. The total duration of a single deep eclipse is approximately 5--6 years.
    
    \item For the first time, we report the discovery of deep periodic photometric minima with a period of $P = 3759 \pm 35$ days. These variations can be attributed to the presence of a close stellar companion or massive planets with a semi-major axis of $a \geq 7.5$ AU. The previously known visual component AS~442B cannot be the cause of the detected periodic change.
    
    \item Our spectroscopic monitoring was executed one year before the deep eclipse of 2023. During 2020 and 2022, we detected a gradual decline in the mean stellar brightness of approximately 0.5 mag. This photometric variability was the primary driver of the observed changes in the spectrophotometric parameters of the emission and absorption lines. Specifically, as the continuum level decreased, we detected a concomitant increase in the intensities of the hydrogen and metal lines. This behavior of spectral variability does not contradict our assumption about the existence of an invisible low-mass companion in the binary system AS~442A.
\end{itemize}

\normalem
\begin{acknowledgements}
The authors thank the AAVSO International Database for providing the photometric observational data used in this study. The authors are also grateful to the Strasbourg CDS Young Star Photometric Data Archive for providing the observational data and to the Simbad and Vizier archive databases. 

The authors are also very grateful to Dr. Mendigutia I. for his very useful advice and comments on this manuscript.

\end{acknowledgements}


\begin{appendix}
\onecolumn
\section{Some figures on spectral variability of AS 442A}

\begin{figure}[ht!]
\centering
\includegraphics[angle=0,width=16cm]{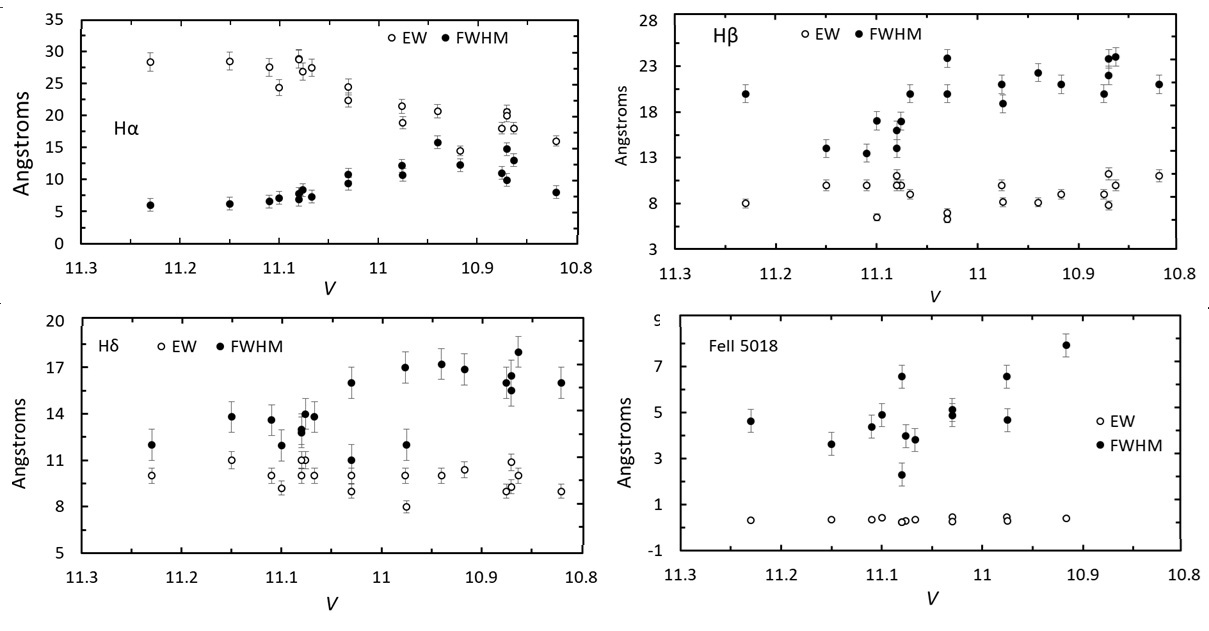}
\caption{The relation between $EW$ and $FWHM$ of the spectral lines H$\alpha$, H$\beta$, H$\delta$, and Fe II 5018 versus the $V$ magnitude.}
\label{Fig1A}
\end{figure}

\begin{figure}[ht!]
\centering
\includegraphics[angle=0,width=12cm]{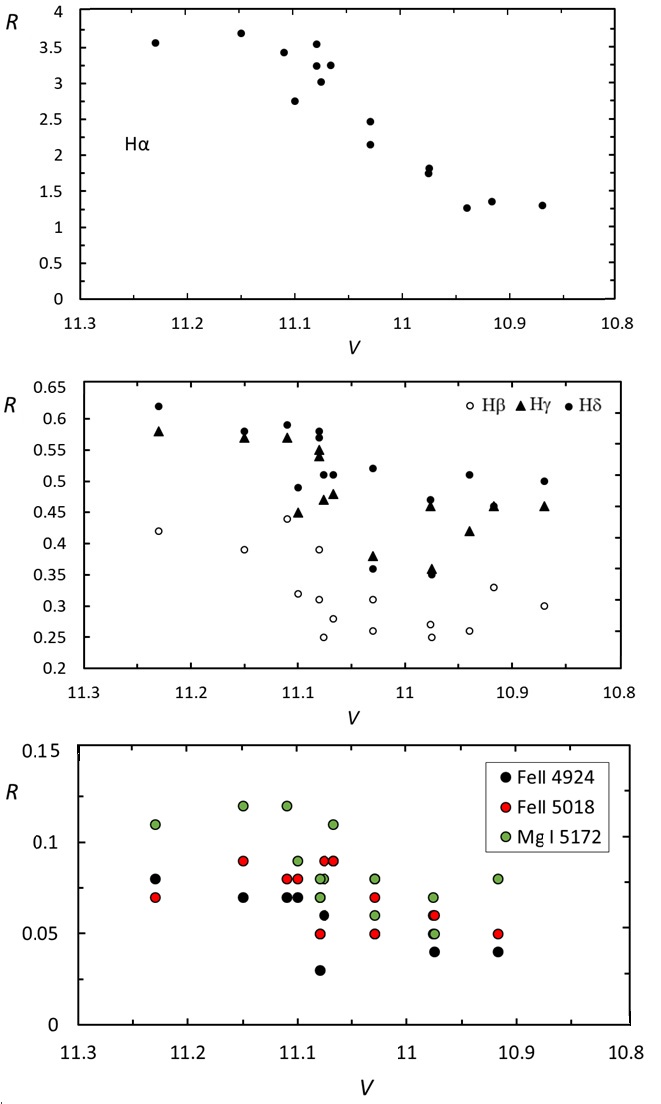}
\caption{The relation between line depth $R$ of the spectral lines H$\alpha$--H$\delta$, and Fe II 5018, 4924, and Mg I 5172 versus the $V$ magnitude.}
\label{Fig2A}
\end{figure}

\begin{figure}[ht!]
\centering
\includegraphics[angle=0,width=16cm]{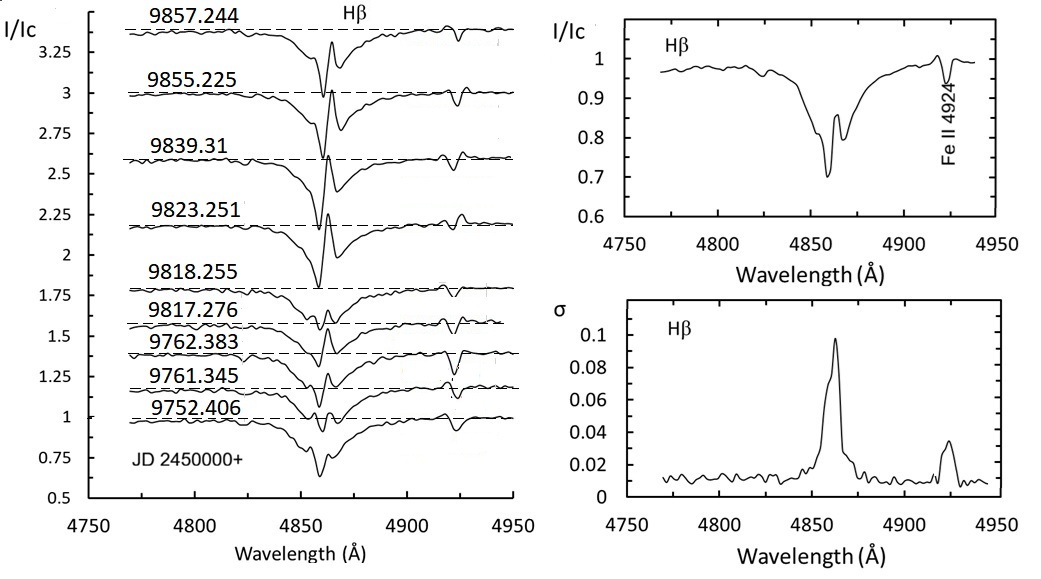}
\caption{The H$\beta$ line profiles obtained in 2022 (left panel). In the right top panel: mean profile of the H$\beta$ line of all observations. In the bottom panel: standard deviation $\sigma$ of the intensity of H$\beta$ line profiles.}
\label{Fig3A}
\end{figure}

\clearpage
\end{appendix}
\end{document}